\documentclass[useAMS,usenatbib,usegraphicx]{mn2e}
\usepackage{inputenc}
\title[Photo-physics of acetic acid and its isomers]
{Large prebiotic molecules in space: photo-physics of acetic
acid and its isomers}
\author[F. Puletti et al.]{Fabrizio Puletti$^1$, Giuliano Malloci$^2$, Giacomo 
Mulas$^2$ and Cesare Cecchi-Pestellini$^2$\thanks{ccp@ca.astro.it}\\
$^1$ Dipartimento di Fisica, Universit\'{a}~ di Cagliari,
Strada Prov.~le Monserrato \textendash Sestu Km~0.700,
I\textendash09042 Monserrato (CA), Italy\\
$^2$ INAF-Osservatorio Astronomico di Cagliari, St. 54 Loc. Poggio dei Pini,
I\textendash09012 Capoterra (CA), Italy}
\include{aas_macros}
\usepackage{graphicx}
\usepackage{times}
\usepackage{amsmath, amscd, amsfonts, amstext, amssymb}
\usepackage{multirow}
\usepackage{verbatim}

\begin{document}

\date{Accepted . Received; in original form }

\maketitle

\begin{abstract}
 An increasing number of large molecules have been positively identified
in space. Many of these molecules are of biological interest and thus
provide insight into prebiotic organic chemistry in the protoplanetary
nebula. Among these molecules, acetic acid is of particular importance 
due to its structural proximity to glycine, the simplest amino acid. 
We compute electronic and vibrational properties of acetic acid and its 
isomers, methyl formate and glycolaldehyde, using density functional theory. 
From computed photo-absorption cross-sections, we obtain the corresponding 
photo-absorption rates for solar radiation at 1 AU and find them in 
good agreement with previous estimates. We also discuss glycolaldehyde 
diffuse emission in Sgr~B2(N), as opposite to emissions from methyl 
formate and acetic acid that appear to be concentrate in the compact 
region Sgr B2(N-LMH).
\end{abstract}


\section{Introduction}
The richness of interstellar chemistry has been growing steadily with
the variety of objects and regions observed. The excitation and abundances
of molecules contain key information on the physical structure and evolution
of the host regions. Through molecules, we can trace the cycle of matter for
interstellar space into stars and planets, and back again into the
interstellar space \citep{HW98}.

There is an increasing body of evidence for the existence of large molecules
in the interstellar medium and in the interplanetary space. Interferometric 
observations of high-mass star forming regions in molecular clouds have 
revealed hot molecular cores, short-lived remnants of clouds not incorporated 
into the newly born massive stars. These hot cores contain within them the 
evaporated material of ices deposited on dust grain surfaces during the 
collapse, and observations show a very interesting chemistry. In particular, 
there exist substantial column densities of large partly hydrogen-saturated 
molecules, many of them being of pre-biotic interest. Some of these species
have been also observed in comets and embedded in minor bodies of the solar 
system \citep{C05}. The study of biologically interesting large interstellar 
molecules offers the exciting opportunity of learning more about the chemical 
evolution preceding the onset of life on the early Earth 4.5 billion years ago.
Comets may be important carriers of prebiotic chemistry, and relevant agents in
the delivery of complex organics to early Earth, as well as to newly formed 
planets.

The Sgr B2 molecular cloud complex is the prime target in the search for
complex species, in particular in a hot core, the so-called Large Molecule
Heimat, Sgr B2(N-LMH), within the more extended molecular cloud. In this
compact source, smaller than the Oort cloud ($\sim0.08$~pc) with a mass of
several thousands M$_\odot$ \citep{MS97}, an extraordinary number of complex
organics have been observed to exhibit very high column densities (e.g. amino
acetonitrile, \citealt{B08}). Large partly hydrogen-saturated species challenge
the completeness of the standard ion-neutral scheme in interstellar chemistry,
suggesting that reactions on dust grains are involved in their formation (e.g.
\citealt{BK07}).

Of the chemical species detected so far, particular attention has been paid to
the formation of different isomer groups. In this work, we focus on
C$_2$H$_4$O$_2$, i.e. acetic acid (CH$_3$COOH), glycolaldehyde (HCOCH$_{2}$OH), and 
methyl formate (HCOOCH$_3$), because of their potential role 
in the origin of life \citep[e.g.,][]{W00,C05}. Glycolaldehyde, the simplest of the 
monosaccharide sugars, has first been detected by \citet{H00} towards Sgr 
B2(N-LMH), and its most recently determined column density in that source is 
$5.9 \times 10^{13}$~cm$^{-2}$ \citep{H06}. It has been recently observed 
outside the galactic center by \citet{B09} towards the hot molecular core 
G31.41+0.31, with the emission coming from the hot and dense region closest to 
the protostars. In addition, \citet{C04} and \citet{D05} presented an upper limit for 
glycolaldehyde abundance in the comet Hale-Bopp. Acetic acid shares the 
C$-$C$-$O backbone with glycine, from which it differs by an amino group 
(NH$_2$). First detected by \cite{M97} in Sgr B2(N-LMH), acetic acid shows a 
column density of $6.1 \times 10^{15}$~cm$^{-2}$ in that region \citep{R02}. 
Finally, methyl formate, discovered towards Sgr B2(N) by \citet{B75}, has been
also observed in other star forming regions of both high \citep{Mc96,G00} and 
low \citep{RH06} mass, towards a proto-planetary nebula \citep{R05}, and in 
comets \citep{BM00,D05,R06}. The column density of methyl formate in  Sgr 
B2(N-LMH) is  $1.1 \times 10^{17}$~cm$^{-2}$ \citep{L01}.

Sgr B2(N-LMH) is the only source where all three of these isomers have been
observed. However, while acetic acid and methyl formate are concentrate in Sgr
B2(N-LMH), glycolaldehyde appears to be more diffusely spread through Sgr
B2(N) \citep{H01}. This behaviour is generally shared by other aldehydes
\citep{S06}. In comets only methyl formate has been observed so far.

An understanding of molecular structure, spectroscopy, and photo-absorption
processes may be of critical importance in interpreting current observations.
In particular, the lifetime of cometary molecules versus photo-destruction
is a basic parameter for all cometary studies: as a matter of fact 
any error in the photodissociation rate translates linearly into an error 
on the abundance derived in the cometary nucleus. 
%in interpreting or predicting observed abundances.
It is also needed for chemical modeling of planetary
atmospheres. In this work, we derive electronic and vibrational properties of 
the isomer triplet C$_2$H$_4$O$_2$ described above using the Density Functional 
Theory (DFT). Photo-destruction rates for acetic acid and methyl formate were
derived by \citet{C94} using old laboratory absorption data published by 
\citet{S88}, while the rate for glycolaldehyde is just an estimate. In
section~\ref{theory} we present a brief outline of the method, together with a
description of computational settings and results. Section~\ref{comets} 
contains the application of these results to cometary photochemistry, while
discussion and conclusions are in the last section.

\section{Theory and results}\label{theory}
We used DFT \citep[e.g.,][]{jon} for the calculation of the equilibrium 
geometry of the electronic ground state and of the vibrational spectrum. We 
then applied the time-dependent extension of the theory \citep[TD-DFT,][]
{marques} to compute the electronic excited states and the resulting 
photo\textendash absorption spectrum for each molecule.
% We performed all electronic ground-state
%calculations using the quantum chemistry program package \textsc{turbomole} 
%\citep{turbomole}.

To obtain the ground\textendash state optimised geometries we used the 
quantum chemistry program package \textsc{turbomole} \citep{turbomole}. 
Technical details about the specific choice of density functional and 
atomic basis set can be found in the Appendix. 

After geometry optimisation, we performed the vibrational analysis
obtaining energies and intensities of the normal modes of vibration in the
harmonic approximation. Vibrational transitions for fundamental configurations 
of the isomer triplet are given in the Appendix.

Finally, keeping fixed the ground state geometries obtained above, we
computed the photo\textendash absorption cross\textendash section for each 
molecule. We used two different implementations of TD\textendash DFT in the 
linear response regime, in conjunction with different representations of 
the Kohn\textendash Sham wavefunctions:
\begin{enumerate}
\item[\it(i)] \label{one}
the real\textendash time propagation scheme using a grid in real space 
\cite{yab99}, as implemented in the \textsc{octopus} computer 
program \citep{mar03};
\item[\it (ii)] \label{two}
the frequency\textendash space implementation \citep{bau96} based on the 
linear combination of localised orbitals, as given in \textsc{turbomole}.
\end{enumerate}

\begin{figure}
\includegraphics{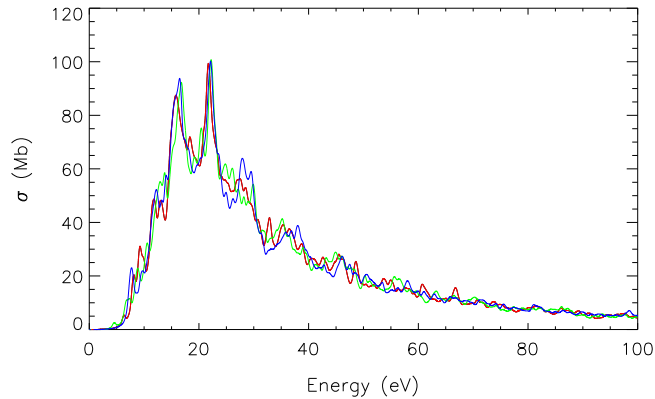}
\caption{Comparison between the photo\textendash absorption 
cross\textendash sections (in Megabarns, 1~Mb = $10^{-18}$~cm$^2$) of acetic 
acid (red), methyl formate (blue), and glycolaldehyde (green), as computed 
up to 100~eV using the real\textendash time TD\textendash DFT implementation 
in the \textsc{octopus} program.}
\label{fig1}
\end{figure}

In the real\textendash time propagation scheme {\it (i)} the whole 
photo\textendash absorption cross\textendash section of the molecule, up to the
far\textendash UV, is obtained at once, which is particularly { convenient} 
for astrophysical applications. Technical details about the TD\textendash DFT 
formalism implemented in the \textsc{octopus} program are reported in the 
Appendix; the resulting spectra are displayed in Fig.~\ref{fig1}. 

\begin{figure}
\includegraphics{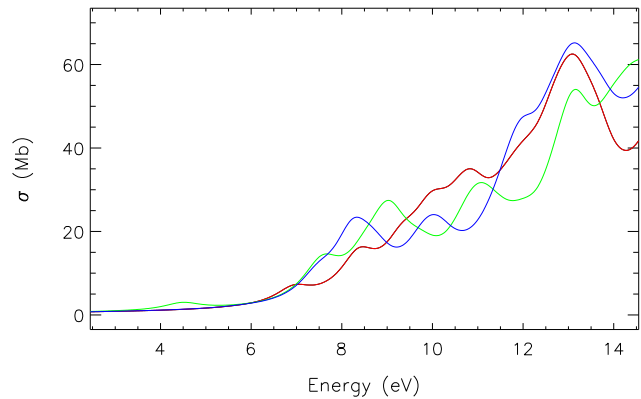}
\caption{Same as Fig.~\ref{fig1} as obtained for the low\textendash energy 
part of the spectrum using the frequency\textendash space TD\textendash DFT 
implementation of the \textsc{turbomole} package at the { BP}/TZVP level 
of theory.}
\label{fig2}
\end{figure}

In the most widely used frequency\textendash space implementation {\it (ii)} the
poles of the linear response function correspond to vertical excitation 
energies and the pole strengths to the corresponding oscillator strengths. 
With this method computational costs scale steeply with the number of required 
transitions; electronic excitations are thus usually limited to the 
low\textendash energy part of the spectrum. Table~\ref{poba} shows that
both combinations BP\textendash TZVP and B3LYP\textendash TZVP 
(see the Appendix for the nomenclature) provide a similarly good
agreement between the computed transitions and the experimental results 
available for acetic acid at room temperature \citep{limao}, 
while BP\textendash TZVP results in a much closer agreement with the experimental 
results for glycolaldehyde 
\citep{kar07}. We therefore chose to use the BP functional for all of our 
calculations.
The resulting BP\textendash TZVP absorption spectra for the three molecules,
shown in Fig.~\ref{fig2}, are obtained as a superposition of Gaussian functions
with fixed arbitrary widths of 0.8~eV. The kind of calculations we 
performed only yield the positions and intensities of vertical, pure electronic
transitions, and therefore give no information on band widths. However, 
in the available gas\textendash phase spectra of acetic acid \citep{limao} and 
glycolaldehyde \citep{kar07} bands up to $\sim 10$~eV show broad profiles with 
a full width at half maximum of about 0.8~eV, produced by the convolution of 
unresolved vibronic structure and the natural width of the transitions.
The list of first 50 excited 
states and transition intensities is given in the Appendix. 

The two TD\textendash DFT implementations produce compatible results in 
the low\textendash energy region, i.~e. up to about 10~eV, while they tend
to diverge significantly at higher energies. However, TD\textendash DFT, as a 
method, is known to yield dependable results for individual transitions 
only for excitation energies up to the ionisation energy, which indeed
is close to $\sim 10$~eV for all three isomers. 

The real\textendash time real\textendash space implementation, on the other 
hand, has been demonstrated to yield good results for the overall density of 
electronic transitions even at high energies, with the \emph{caveat} that 
single peaks of fine structure are meaningless 
\citep[see e.~g.][]{mar03}: at energies $\gtrsim 10$~eV only the 
\emph{envelope} of the spectra calculated by octopus is expected to be 
accurate, while the resolved fine structure is largely due to standing waves 
in the finite simulation box, is strongly dependent on the size of the box and
is only partly quenched by the absorbing boundary conditions we adopted. 
Since we use these spectra at high energies to compute absorption rates in a 
continuous spectrum (see Sect.~\ref{comets}), the effect of spurious fine 
structure averages out when integrating over ranges of many eVs, 
making these theoretical spectra quite adequate for their intended purpose.

\begin{table}
\caption{Comparison between the computed electronic transitions 
as obtained using different exchange\textendash correlation functionals and 
the experimental data reported by \citet{limao} for acetic acid and 
by \citet{kar07} for glycolaldehyde.}
\label{poba}
\begin{center}
\begin{tabular}{ccc}
\hline \hline
B-P/TZVP & B3LYP/TZVP & EXP\\
\hline
\multicolumn{3}{c}{acetic acid} \\[4pt]
5.60 & 5.84 & 6.09 \\
6.93 & 7.49 & 7.22 \\
8.20 & 8.30 &  8.15 \\
8.25 & 8.50 & 8.35 \\
9.20 & 9.25 & 8.82 \\
10.07 & 10.54 & 10.29 \\[4pt]
\multicolumn{3}{c}{Average relative error (\%)}\\
3.4 & 3.2 & \\[4pt]
\hline
\multicolumn{3}{c}{glycolaldehyde} \\[4pt]
4.49 & 5.61 & 4.51 \\[4pt]
\multicolumn{3}{c}{Average relative error (\%)}\\
0.4 & 24.4 & \\
\hline
\end{tabular}
\end{center}
\end{table}

\section{Photochemistry} \label{comets}
Photo-absorption rates of the three isomers are computed by means of the
relation
\begin{equation}
\beta \, ({\rm s}^{-1})= \int_{\Delta E} S_\odot (E) \sigma(E) {\rm d} E 
\end{equation}
where, for the sake of comparison, $S_\odot (E)$ is the solar spectrum
at 1 AU provided by \citet{H92}, and $\sigma (E)$ the 
photo\textendash absorption cross\textendash section for a given molecule. 
The resulting photo\textendash absorption rates are 
reported in Table~\ref{sole}, where we also show results for a radiation
density expected in a photodissociation front near an OB star \citep{draine}. 
We also list separately the contribution of the Ly$\alpha$ line in the 
solar spectrum to the absorption rates, which is of the order of $\sim$15\% 
of the total, in agreement with previous estimates of \citet{C94}.
As evident in Fig.~\ref{fig2} and Table~\ref{stigazzi}, glycolaldehyde 
presents a relatively strong band near 4.5~eV, at an energy $\sim$2~eV lower 
than the first transitions of comparable intensity in the other two members of 
the triplet (namely $\sim$6.9~eV for acetic acid and $\sim$7.5~eV for methyl 
formate). Since the solar spectral distribution decreases very steeply in this 
energy range, the estimated glycolaldehyde photo\textendash absorption rate 
in the solar radiation field is by and large dominated by this single band, 
and results to be more than 100 times larger than those of acetic acid 
and methyl formate. This effect is not present for 
photo\textendash absorption rates estimated for the \citet{draine} spectrum, 
since it is much flatter than the solar one in the UV, and thus no single band 
dominates the absorption for any of the three species.
TD\textendash DFT calculations are known to provide transition energies 
accurate within about 0.3~eV. 
To test the robustness of our results we shifted the energies of all 
transitions by $\pm$0.3~eV, and checked how this affects the resulting 
photo\textendash absorption rates. Also, since 
TD\textendash DFT provides no information on the intrinsic width of the 
calculated bands, we assumed for all bands a full\textendash width at
half maximum of $\sim$0.8~eV, consistent with published experimental data 
\citep{limao,kar07}. The overall variations in photo\textendash absorption 
rates are within a factor of 2, meaning that our conclusions are rather firm.
This accuracy may appear surprising for the estimated absorption rates in 
the Ly$\alpha$ line, which would naively be expected to vary very strongly 
with the calculated positions of molecular bands, producing either a very 
large absorption rate if it happens to be close to a strong one or a very 
small one if it falls in a gap in the absorption spectrum; however, the 
assumption (based on experimental data) of a 0.8~eV FWHM for all bands vastly
reduces the dependence of absorption rates on small (i.~e. $\sim$0.3~eV) 
variations in band positions.

\begin{table*}
\caption{Photo-absorption rates (s$^{-1}$) for the isomer triplet 
C$_2$H$_4$O$_2$ in two radiation fields, assuming a band FWHM of 
$\sim$0.8~eV based on available experimental data \citep{limao,kar07}.
The first column lists the computed absorption rates for the whole 
solar spectrum at 1~AU \citep{H92}, the second column
in the solar Ly$\alpha$ line at 1~AU \citep{H92}, the third in a 
photodissociation front near an OB star \citep{draine}.}
\label{sole}
\begin{center}
\begin{tabular}{llccc}
\hline \hline
\multicolumn{2}{c}{Species} & Solar flux at 1 AU & Solar Ly$\alpha$ at 1 AU & 
Unshielded OB field \\
& & \multicolumn{2}{c}{\citep{H92}} & \citep{draine}\\[2pt]
\cline{3-5} 
& & & \\
Acetic acid & CH$_3$COOH & \phantom{$^{(1)}$}6.7(-5)$^{(1)}$ & 9.7(-6) & 
2.9(-9)\\
Glycolaldehyde$^{(2)}$ & HCOCH$_{2}$OH & 9.6(-3) & \textemdash& 1.3(-10) \\
& & 2.8(-5) & 3.7(-6) & 2.9(-9) \\
Methyl formate & HCOOCH$_3$ & 5.0(-5) & 7.4(-6) & 3.0(-9) \\
\hline \hline
\end{tabular}
\flushleft
\flushleft
$^{(1)}$ $1.8(-5) = 1.8 \times 10^{-5}$.
\flushleft 
$^{(2)}$ First row: photo\textendash absorption rates due to the first two 
absorption bands (up to $\sim$4.5~eV); second row: photo\textendash absorption 
rates due to the remaining bands.
\end{center}
\end{table*}

In a recent analysis of ice composition in comet Hale\textendash Bopp, 
\citet{C04} quote a photodestruction rate (at 1 AU) for acetic acid 
$\beta_{\rm CH_3COOH} = 5.1 \times 10^{-5}$ s$^{-1}$. This value was provided 
by \citet{C94}, that also reported a rate 
$\beta_{\rm CH_3OHCO} = 4.7 \times 10^{-5}$ s$^{-1}$ for methyl formate. 
The photodestruction rate of glycolaldehyde is unknown and assumed to be 
$1 \times 10^{-4}$ s$^{-1}$ at 1 AU \citep{C04}. Assuming a unit 
photodestruction yield and a factor of 2 indetermination in the calculations, 
our photodestruction rates are consistent with \citet{C94} values, that were
based on old laboratory data provided by \citet{S88}. The case of 
glycolaldehyde is different, since most of photo\textendash absorption is 
produced by the band at 4.5~eV, that appears to be too low in energy to 
provide a unit photodestruction yield. 
%{ Inserire commenti sul perche' pensiamo che sia cosi'. [GMu Giulio o 
%Fabrizio, potreste fare una ricerca bibliografica per trovare anche solo un 
%papero che mostri l'andamento dello yield di fotodistruzione di una molecola 
%simile per gruppi funzionali e (grosso modo) per numero di atomi? Se mi date 
%quello, a scrivere il testo conseguente ci penso io.] } 
As a consequence, photodestruction channels are 
activated just via absorption in the high\textendash energy bands, leading to 
a photo\textendash absorption rate 
$\beta^\star_{\rm CH_2OHCHO} \sim 3 \times 10^{-5}$ s$^{-1}$ (cf. Table 
\ref{sole}), much lower than the rate assumed in \citet{C04}. 

\begin{figure}
\includegraphics[width=\hsize]{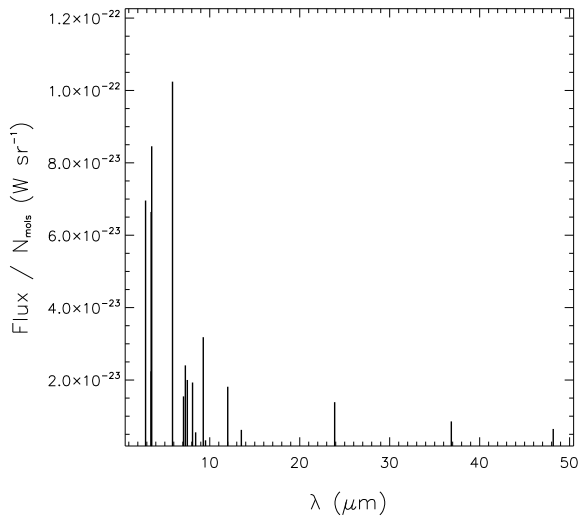}
\caption{IR emission spectrum of one glycolaldehyde molecule at 1~AU from
the quiet sun, estimated by assuming that all the energy absorbed in 
electronic transitions is converted in vibrational excitation and subsequently
reradiated (see text). Emission in each vibrational band is represented by a 
vertical bar whose abscissa corresponds to the wavelength of the vibrational 
mode and whose height equals the calculated emission intensity.}
\label{irsugar}
\end{figure}

Photons absorbed in the lower energy bands of CH$_2$OHCHO may be resonantly
scattered, or, if the molecule undergoes internal conversion, 
re\textendash emitted in the IR range. Using an adapted version of a 
Monte~Carlo model developed for emission by polycyclyc aromatic hydrocarbons, 
which is indeed assumed to be pumped by complete internal conversion of the 
energy absorbed via electronic transitions in the visible and UV 
\citep{mulas}, we have constructed the expected IR emission by glycolaldehyde, 
powered by the solar flux at 1 AU, assuming all the energy absorbed in the 
band at $\sim$4.5~eV to be emitted in the IR. The emission coefficient 
(per molecule) is reported in Fig. \ref{irsugar}. 

We have thus far considered photo\textendash absorption rates in the standard 
solar radiation field at 1~AU (for the ``quiet'' Sun), which are relevant for 
the photochemistry of glycolaldehyde, acetic acid and methyl formate in our
present solar system. However, from the study of
stellar proxies for the Sun it appears that young solar type stars emit high
energy photons at a level three to four orders of magnitude higher than the
present-day Sun, both during the pre-main sequence phase when the emission
is dominated by intense daily or weekly flares \citep{Fav05}, and
during the first phases of the main sequence \citep{Mic02}.
Therefore, chemical evolution can only be understood within the context of
the evolving stellar radiation environment.

\begin{figure}
\includegraphics[angle=270,width=\hsize]{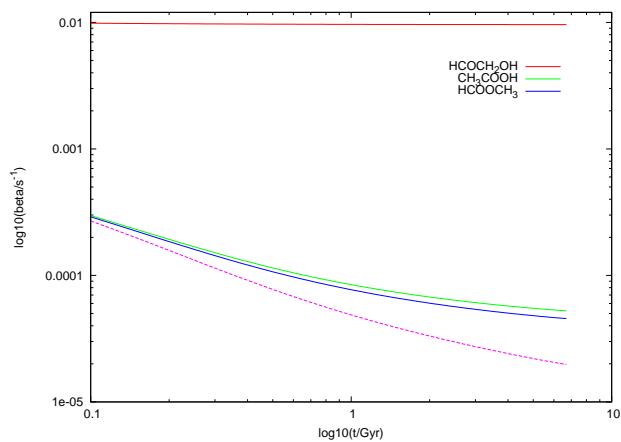}
\caption{Photo-absorption rates in the environments of solar-type stars
of different ages. Top to bottom: glycolaldehyde, acetic acid, methyl formate,
and glycolaldehyde with the band at 4.5~eV supressed.}
\label{evol}
\end{figure}

Without addressing the problem of molecular survival in a disk 
\citep[e.~g.][]{Vis07}, we estimate the effect of the extreme UV emission 
(roughly the spectral range 
between 13 and 100 eV) from solar\textendash type stars of different ages,  
exploiting the emission of six stars from the Sun in Time program 
\citep{GR02}, 
whose fluxes are assumed to describe the evolution of the Sun's emission 
\citep{Ri05}.
Since \citet{Ri05} reported integrated fluxes (at 1 AU), the rate coefficents 
are approximated by 
\begin{equation}
\beta(t) \, ({\rm s}^{-1}) = \sum_{\Delta E} \langle \sigma \rangle_{\Delta E} 
~ \frac { S_\odot (t,\Delta E) } { \Delta E }
\end{equation} 
where $\langle \sigma \rangle_{\Delta E}$ is the average over the energy range
$\Delta E$ of the photo\textendash absorption cross\textendash sections shown 
in Fig. \ref{fig1}.
$\Delta E$ intervals are taken from \citet{Ri05}, Table 4. Beyond 100 eV a
collective description of the molecule is not anymore necessary, since X-ray 
absorption cross\textendash sections can be closely approximated by adding 
the atomic cross\textendash sections of individual atoms bound in the 
molecule (e.g., \citealt{CCP06}). We therefore do not perform 
calculations in the X-ray energy range here. 
Results are reported in Fig.~\ref{evol}. It is evident that the high
energy tail of stellar spectrum provides a significant enhancing of 
photo\textendash absorption rates for acetic acid and methyl formate. 
Glycolaldehyde photon absorption rates are not changing too much, although 
the extreme UV contribution to photo\textendash absorption is comparable to 
absorption in the near UV band at $\sim$4.5~eV. In general, due to the 
increase in the stellar high energy component, the 
photo\textendash destruction rates of the isomer triplet members increase 
roughly two order of magnitude in the environment of a young 
solar\textendash like star.
 
\section{Discussion and conclusions}
In this work we studied the photo\textendash physics of the acetic acid and 
its isomers, glycolaldehyde and methyl formate. Computed 
photo\textendash absorption rates are consistent with literature data for 
acetic acid and methyl formate.
In the case of glycolaldehyde, for which the photo-absorption was completely 
missing up to the far\textendash UV, our calculations indicate that 
photodestruction is slower than in the other two members of the isomer 
triplet. However, the overall photo\textendash absorption rate is much 
larger, and likely to produce either 
resonant scattering or IR emission powered by near UV solar photons.  

Sgr B2 observations of glycolaldehyde show that, unlike acetic acid and methyl
formate, its emission is extended in the surrounding molecular cloud. Such a 
behaviour, that has not been understood so far (e.g. \citealt{chenga}), is 
part of a more general problem involving differentiation in isomers, such as 
e.g., isocyanide isomers (CH$_3$CN, CH$_3$NC). The behaviour of
this isomer triplet has also recently been discussed by \citet{lat09} as a
notable exception to what was called the ``Minimum Energy Principle'' (MEP) 
whereby whenever several isomers are possible for a given formula, their 
observed abundances are in order of binding energy. The authors made the 
educated guess that this anomaly is first created by differences in the 
chemical pathways leading to the formation of glycolhaldehyde, acetic acid 
and methyl formate on the surfaces of dust grains, 
and then preserved due to large energy barriers for the 
conversion among them. 
Our results may help to shed some more light on this problem. Although
both hot core and the embedding molecular cloud are dark regions, 
cosmic\textendash rays provide a source of UV photons at high visual 
extinctions by exciting molecular hydrogen in the Lyman and Werner bands 
\citep{taraf}. This locally generated photon flux typically has fluences 
lower than $10.000$ photons cm$^{-2}$ s$^{-1}$ \citep{ccp92}, and may produce 
important chemical effects \citep{gredel,BK07}. However, the similarity in 
both the photo\textendash absorption cross\textendash sections and 
ionisation potentials of the three species makes chemical differentiation 
due to selective photo\textendash destruction unlikely.
As a possible explanation of the extended spatial scale of 
glycolhaldehyde, we consider the possibility of a slow, selective 
isomerisation, i.e. the possibility 
that a species may convert itself into another member of the C$_2$H$_4$O$_2$ 
triplet by interacting with the radiation field. This would imply 
an isomerisation mechanism which operates on a timescale which is much longer
than the typical lifetime of a hot core 
\citep[few times $10^4$~years,][]{wilner}
but still short enough to be effective on the timescale of the lifetime of a
molecular cloud \citep[$\sim10^8$~years,][]{blitz00}. In this framework, hot 
cores would
reflect the relative abundances among the isomers as created by their 
production mechanisms, whereas in the molecular cloud the isomers would 
``relax'' to the most stable one, namely glycolaldehyde, in agreement with
the MEP.

Isomerisation may be induced by the absorption of radiation essentially in 
two main ways; following the absorption of a UV\textendash visible photon 
the molecule could
\begin{enumerate}
\item[\it (i)]
move to an electronic state whose energy surface presents a minimum 
close to the equilibrium configuration of another isomer;
\item[\it (ii)]
convert, via one or more non\textendash radiative transitions, a substantial 
part of the electronic excitation energy into  vibrational energy allowing the 
overcoming of the isomerisation potential barriers.
\end{enumerate}
The present data do not allow a discrimination between the two cases. Moreover,
the analysis of the first isomerisation channel would require a detailed study 
of the energy hypersurfaces in the excited states accessible with photons 
generated by the \citet{taraf} mechanism. Therefore, we will discuss 
qualitatively the second isomerisation process. We assume that every absorption
heats up the molecule in a time scale characteristic of electronic transitions
($\sim 10^{-8}$~s), that then decades via non\textendash radiative transitions 
($\lesssim 10^{-10}$~s). The electronic excitation energy can be uniquely 
released by a cascade of vibrational transitions. We also assume, as 
simplifying hypotheses, that (\textit{i}) all the excitation energy is 
converted in vibrational excitation, (\textit{ii}) that there exist only one 
isomerisation barrier, (\textit{iii}) that the excitation energy is far greater 
than this barrier and (\textit{iv}) that the isomerisation rate is high above 
the threshold and zero below it. Then every time a molecule absorbs an 
UV\textendash visible photon, it will be vibrationally heated. If the energy 
of the absorbed photon is above the isomerisation threshold, the molecule 
will establish a statistical equilibrium, in which the probability to find it 
in one of the isomeric configurations will be proportional to the density of 
vibrational states at the given energy. The molecule will then cool down, in 
timescales of the order of a second, with a vibrational cascade (all 
vibrational modes are IR\textendash active for all three isomers). Its energy will thus 
eventually fall below the 
barrier for the isomerisation. When this happens, the proportion among the 
isomers is frozen because the conversion rate among the species drops to zero, 
the abundances are therefore those given by the ratios among the 
densities of vibrational states. Since we are here dealing with three 
conformations, one should consider at least three different isomerisation 
channels, each with its different barrier(s). Whatever they are, however, 
as long as they are easily overcome with the energy of a single UV photon, 
the ratio of the abundances of the isomers, if they are exposed to UV light, 
should be fixed at the ratios of the densities of states at the threshold(s).
\begin{figure}
\includegraphics[width=\hsize]{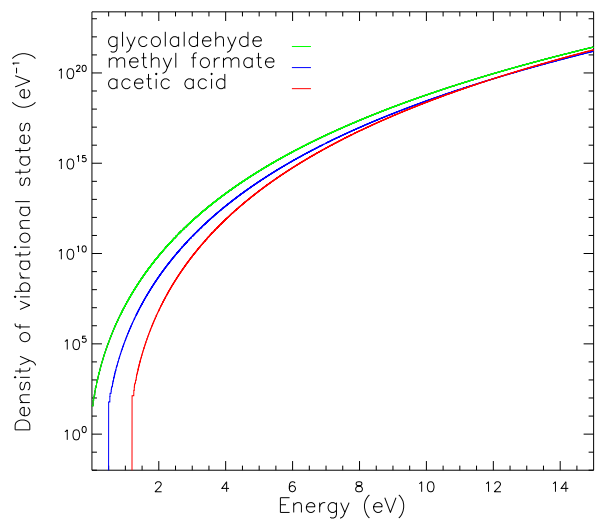}
\caption{Densities of the vibrational states for the three species as
a function of energy.\label{dos}}
\end{figure}

\begin{table}
\caption{Total energies for the three species considered. 
The second column shows the energy difference $\delta$ in comparison with 
glycolaldehyde.}
\label{tab:rate}
\begin{center}
 \begin{tabular}{ccc}
\hline \hline
 & Total energy (eV) & $\delta$ (eV)\\
\hline
Glycolaldehyde & -6231.83 & /\\
Metyl formate & -6232.32 &  0.49\\
Acetic acid & -6233.04 & 1.21\\
\hline \hline
 \end{tabular}
\end{center}
\end{table}

The densities of vibrational states for the three species, calculated in 
harmonic approximation through the algorithm of \cite{stein}, is shown in 
Fig.~\ref{dos}. The distances among the curves are due to the 
differences in the total energies shown in Table~\ref{tab:rate} and 
the differences in the frequencies of the vibrational modes. 

The naive model drawn above would essentially turn all isomers into
glycolaldehyde as soon as they absorb an UV photon. However, UV 
absorption, in the weak radiation field produced by to cosmic\textendash 
ray induced H$_2$ fluorescence, occurs on timescales of the order of
$\ga 50.000$ years, that are comparable to the lifetime of hot cores.
As consequence, MEP should not operate effectively in the hot core phase.

Therefore, one of the simplistic assumption in the naive model must be 
incorrect, slowing down isomerisation. Either internal conversion has a low
quantum yield for these molecules, meaning that they get vibrationally heated
only a small fraction of the times they absorb an UV photon; or isomerisation 
is \emph{not} fast whenever vibrational excitation is sufficient to overcome 
reaction barriers, due to the morphology of the molecular potential energy
surface. The latter can easily be the case if the unimolecular reaction
path(s) to isomerisation are narrow and complicated in terms of phase space of
the ions.

Of course, conversion among isomers needs not to proceed \emph{only} via 
unimolecular reactions induced by radiation: any conversion reaction including
chemical reactions with abundant enough partners and without activation 
barriers could equally well operate on the right timescales to fulfil the MEP
in molecular clouds but not in hot cores.
A thorough study of the potential energy surface of these three isomers, 
including the reaction paths connecting them, is called for in order to further
progress in the understanding of this observational puzzle.

\section*{Acknowledgments}
We acknowledge financial support by MIUR under project CyberSar, 
call 1575/2004 of PON 2000\textendash2006. We thank both referees for 
their useful comments and suggestions about significant literature 
on this subject which we had previously missed.

\section*{Appendix}
\label{appendix}

We performed all calculations in the framework of Density Functional Theory, 
using the quantum chemistry packages \textsc{turbomole} \citep{turbomole} and
\textsc{octopus} \citep{mar03}.

\subsection{Geometry optimizations}

%Geometry configurations have been obtained using the quantum 
%chemistry program \textsc{turbomole} \citep{turbomole}. 

To find the electronic ground\textendash state geometry of the three molecules 
considered, we first tested different combinations of density functional and 
basis set exploiting \textsc{turbomole}, to identify the most suitable one for 
our purposes. To this aim, we compared the 
experimental ground state geometry of the acetic acid \citep{limao} with the 
one we obtained using the combinations: BP\textendash SV(P), 
BP\textendash TZVP, B3LYP\textendash SV(P), and 
B3LYP\textendash TZVP (see the \textsc{turbomole} 
manual\footnote{www.turbomole.com} and 
references therein for precise definitions of basis sets and functionals). 
All combinations are in good agreement with experimental and previous 
theoretical data \citep{limao}, with the hybrid functional B3LYP 
showing a very slightly lower relative average error.
This is not unexpected, since this functional is known to produce good
results with other classes of organic molecules \citep[e.g.,][]{mar96}.
Considering that both functionals show the same relative average error
with both basis sets, we chose the larger TZVP basis set,
since this is suggested by the \textsc{turbomole}
manual as the default to get reliable quantitative results.

\subsection{Vibrational properties}

For our modelling purposes, we calculated the harmonic vibrational 
frequencies of acetic acid, glycolaldehyde, and methyl formate. All 
calculations were performed at the BP/\mbox{TZVP} level using the 
\textsc{turbomole} program package, and resulted compatible with 
previously published results \citep{limao,sen04,sen05}.
The absolute intensities $S$ of the 
IR\textendash active modes are given in units of km$\cdot$mol$^{-1}$. For 
each vibrational mode of frequency $\tilde{\nu}$ expressed in cm$^{-1}$ the 
corresponding Einstein $A$ coefficients for spontaneous emission can be 
computed as:
\begin{displaymath}
A\,(\mathrm{s}^{-1}) = 
{8 \,\pi \over \mathrm{N}_\mathrm{A} c}\,\tilde{\nu}^2\,S \simeq 
\end{displaymath}
\begin{displaymath}
\simeq 1.2512 \times 10^{-7}  {\left ( \frac {\tilde{\nu}}{\mathrm{cm}^{-1}} \right )}^2\,
{\left ( \frac {S}{\mathrm{km}\cdot\mathrm{mol}^{-1}} \right )}\,
{\mathrm{mol}\cdot\mathrm{cm}^2\over \mathrm{km}\cdot\mathrm{s}}, 
\end{displaymath}
$\mathrm{N}_\mathrm{A}$ being the Avogadro's constant and $c$ the 
velocity of light.

\subsection{Electronic spectra}

In the real\textendash time implementation of TD\textendash DFT as given in 
\textsc{octopus}, the time\textendash dependent Kohn\textendash Sham 
equations are directly integrated in real time and the wavefunctions are 
represented by their discretised values on a spatial grid. 
The static Kohn\textendash Sham wavefunctions are perturbed by an impulsive 
electric field and propagated for a given finite time interval. In this way, 
all of the frequencies of the system are excited. The whole absolute 
absorption cross\textendash section $\sigma(E)$ then follows from the 
dynamical polarisability $\alpha(E)$, which is related to the Fourier 
transform of the time\textendash dependent dipole moment of the molecule. 
The relation is:
\begin{equation}
\label{sigma}
\sigma(E) = \frac{8 \pi^2 E}{h c}\,\Im \{\alpha(E)\},
\end{equation}
where $h$ is Planck's constant, $\Im \{\alpha(E)\}$ is the imaginary part of
the dynamical polarisability, and $c$ the velocity of light in vacuum. 

We performed the \textsc{octopus} calculations using 
\citet{bec88} exchange and \citet{per86} correlation functionals. 
The ionic potentials are replaced by norm\textendash conserving 
pseudo\textendash potentials \citep{tro91}. We used a grid spacing of 0.12~\AA{}
and a box size of 6.8~\AA, determined by convergence tests on ground state 
properties and on the photo\textendash absorption spectrum at energies $\lesssim 10$~eV. 
This box size ensures that each atom is at least 4~\AA{} away from its edges. 
We furthermore added a 1~\AA{} thick absorbing 
boundary, which partially quenches spurious resonances due to standing waves 
in the finite simulation box used to confine the molecules 
\citep{yab99,mar03}. We used a total time integration length T=20 $\hbar/$eV, 
corresponding to an energy resolution of $\hbar/T$=0.05~eV. 
For the numerical integration of the time evolution we used a time step of 
0.0008~$\hbar/$eV, which ensured energy conservation with good accuracy, 
within numerical noise.

In the most widely used frequency\textendash space TD\textendash DFT 
implementation, the poles of the linear response function correspond to 
vertical excitation energies and the pole strengths to the corresponding 
oscillator strengths. With this method computational costs scale steeply 
with the number of required transitions and electronic excitations are thus 
usually limited to the low\textendash energy part of the spectrum. The 
frequency\textendash space TD\textendash DFT calculations with 
\textsc{turbomole} were performed at the BP/\mbox{TZVP} level of theory,
since this showed the best agreement with experimental data.
We report in following Table~\ref{stigazzi} the first 60 
singlet\textendash singlet electronic transitions of the three molecules 
under study with the corresponding oscillator strengths. 

\begin{table}
\caption{Vertical electronic transitions (eV) and corresponding oscillator
strengths of the three molecules considered, as computed at the 
BP/\mbox{TZVP} level using the \textsc{turbomole} program package.}
\label{stigazzi}
\begin{center}
\begin{tabular}{cccccc}
\hline \hline
\multicolumn{2}{c}{Acetic acid} & \multicolumn{2}{c}{Glycolhaldehyde} & 
\multicolumn{2}{c}{Methyl formate}\\
\hline
5.70 & 7.3(-4) & 4.46 & 3.6(-5) & 5.82 & 1.3(-3) \\
6.94 & 5.4(-2) & 4.49 & 2.4(-2) & 7.46 & 6.9(-2) \\
7.74 & 4.7(-3) & 6.43 & 2.9(-4) & 7.70 & 1.5(-3) \\
8.25 & 2.6(-3) & 7.34 & 3.2(-3) & 7.86 & 3.7(-3) \\
8.37 & 1.3(-1) & 7.35 & 1.1(-2) & 8.24 & 1.9(-1) \\
8.82 & 1.9(-4) & 7.57 & 8.7(-2) & 8.54 & 3.0(-2) \\
9.05 & 3.3(-3) & 7.61 & 3.0(-2) & 8.74 & 7.6(-2) \\
9.08 & 1.4(-4) & 8.48 & 6.1(-2) & 8.83 & 4.3(-3) \\
9.16 & 8.1(-4) & 8.50 & 1.6(-2) & 9.27 & 1.3(-4) \\
9.27 & 2.9(-3) & 8.63 & 1.1(-3) & 9.39 & 7.3(-4) \\
9.33 & 1.2(-1) & 8.83 & 1.7(-2) & 9.48 & 4.2(-5) \\
9.88 & 1.1(-1) & 8.90 & 4.3(-2) & 9.68 & 3.8(-2) \\
10.03 & 6.2(-4) & 9.05 & 2.0(-1) & 9.78 & 1.1(-2) \\
10.07 & 1.1(-1) & 9.51 & 3.0(-2) & 9.89 & 8.1(-2) \\
10.36 & 6.6(-4) & 9.66 & 3.4(-4) & 10.11 & 9.2(-2) \\
10.52 & 5.1(-3) & 9.69 & 6.8(-3) & 10.24 & 2.3(-2) \\
10.58 & 8.8(-2) & 9.77 & 6.0(-2) & 10.36 & 9.4(-4) \\
10.80 & 9.2(-2) & 10.08 & 9.0(-3) & 10.58 & 1.4(-2) \\
10.88 & 3.4(-2) & 10.47 & 4.0(-4) & 10.66 & 7.6(-3) \\
10.92 & 4.8(-2) & 10.52 & 6.0(-3) & 10.96 & 8.6(-3) \\
10.95 & 3.1(-2) & 10.60 & 1.3(-3) & 11.13 & 1.4(-2) \\
11.09 & 1.7(-3) & 10.66 & 6.5(-2) & 11.16 & 1.6(-3) \\
11.44 & 2.1(-2) & 10.70 & 4.6(-3) & 11.24 & 4.4(-4) \\
11.51 & 6.7(-2) & 10.87 & 2.8(-3) & 11.26 & 2.3(-2) \\
11.60 & 7.1(-3) & 10.93 & 1.2(-1) & 11.34 & 1.1(-2) \\
11.75 & 9.7(-3) & 10.95 & 1.2(-3) & 11.46 & 1.3(-2) \\
11.82 & 4.1(-6) & 11.05 & 8.3(-4) & 11.61 & 7.9(-2) \\
11.84 & 1.2(-1) & 11.15 & 1.2(-1) & 11.68 & 2.1(-2) \\
12.09 & 7.9(-2) & 11.46 & 6.2(-2) & 11.84 & 6.3(-2) \\
12.11 & 1.4(-3) & 11.62 & 1.0(-2) & 11.96 & 1.8(-1) \\
12.17 & 4.5(-2) & 11.64 & 3.9(-3) & 11.97 & 4.2(-3) \\
12.21 & 6.0(-4) & 11.95 & 5.4(-2) & 11.99 & 3.6(-2) \\
12.34 & 9.0(-3) & 12.03 & 4.1(-2) & 12.07 & 1.5(-2) \\
12.45 & 1.1(-2) & 12.26 & 1.5(-3) & 12.11 & 9.0(-3) \\
12.58 & 1.1(-1) & 12.37 & 7.9(-4) & 12.26 & 6.5(-3) \\
12.71 & 8.5(-2) & 12.60 & 4.5(-2) & 12.48 & 5.5(-2) \\
12.75 & 4.3(-2) & 12.71 & 2.9(-2) & 12.57 & 1.4(-1) \\
12.88 & 1.4(-2) & 12.72 & 3.0(-3) & 12.70 & 8.0(-3) \\
12.89 & 1.3(-2) & 12.72 & 3.1(-2) & 12.71 & 1.8(-4) \\
12.92 & 4.0(-2) & 12.80 & 2.3(-3) & 12.84 & 1.3(-1) \\
12.93 & 2.4(-2) & 12.87 & 7.5(-3) & 12.92 & 1.1(-2) \\
13.00 & 5.6(-3) & 12.92 & 1.5(-2) & 13.02 & 4.4(-3) \\
13.05 & 1.3(-1) & 12.96 & 8.7(-2) & 13.03 & 1.4(-1) \\
13.10 & 5.6(-2) & 13.11 & 4.3(-2) & 13.16 & 1.6(-1) \\
13.21 & 2.0(-1) & 13.13 & 2.1(-1) & 13.24 & 5.4(-2) \\
13.47 & 2.7(-2) & 13.19 & 5.2(-2) & 13.28 & 1.4(-2) \\
13.53 & 1.3(-2) & 13.23 & 5.8(-2) & 13.42 & 5.3(-2) \\
13.59 & 1.3(-1) & 13.61 & 1.1(-2) & 13.42 & 4.9(-2) \\
13.66 & 4.4(-2) & 13.68 & 3.9(-2) & 13.61 & 1.0(-2) \\
13.76 & 7.9(-2) & 13.72 & 2.8(-2) & 13.66 & 1.4(-1) \\
13.77 & 2.9(-4) & 13.83 & 1.7(-2) & 13.74 & 5.8(-2) \\
14.10 & 6.4(-3) & 13.89 & 2.4(-2) & 13.83 & 2.1(-2) \\
14.22 & 4.0(-2) & 13.94 & 7.8(-2) & 13.91 & 3.3(-2) \\
14.23 & 6.0(-3) & 14.00 & 2.8(-2) & 14.02 & 2.9(-2) \\
14.25 & 9.8(-3) & 14.02 & 4.1(-2) & 14.06 & 3.2(-3) \\
14.41 & 2.2(-4) & 14.07 & 1.4(-2) & 14.14 & 6.3(-3) \\
14.50 & 5.1(-3) & 14.17 & 5.2(-2) & 14.23 & 5.1(-2) \\
14.62 & 3.2(-2) & 14.24 & 5.1(-3) & 14.28 & 2.3(-2) \\
14.65 & 6.2(-2) & 14.39 & 2.0(-1) & 14.29 & 5.1(-2) \\
14.68 & 9.2(-3) & 14.47 & 4.0(-2) & 14.52 & 8.5(-3) \\
\hline \hline
\end{tabular}
\end{center}
\end{table}

\end{document}